# On properties of non-markovian random walk in one dimension

**Manoj C. Warambhe**[1] **and Prashant M. Gade**[1]

[1] Department of Physics Rashtrasant Tukadoji Maharaj
Nagpur University, Nagpur, 440033, India

E-mail: manojwarambhe8@gmail.com

**Abstract.** We study a strongly Non-Markovian variant of random walk in which the probability of visiting a given site i is a function f of number of previous visits $v(i)$ to the site. If the probability is proportional to number of visits to the site, say $f(i)=(v(i)+1)^\alpha$ the probability distribution of visited sites tends to be flat for $\alpha>0$ compared to simple random walk. For $f(i)=\exp(-v(i))$, we observe a distribution with two peaks. The origin is no longer the most probable site. The probability is maximum at site $k(t)$ which increases in time. For $f(i)=\exp(-v(i))$ and for $\alpha>0$ the properties do not change as the walk ages. However, for $\alpha<0$, the properties are similar to simple random walk asymptotically. We study lattice covering time for these functions. The lattice covering time scales as $N^z$, with $z=2$, for $\alpha \leq 0$, $z>2$ for $\alpha >0$ and $z<2$ for $f(i)=\exp(-v(i))$.

**Keywords:** random walk, self-avoiding walk.

## 1. Introduction

The random walk has been one of the most recurring themes in several disciplines in sciences ranging from chemistry and ecology to search algorithms on web and motion of stars in the galaxy [1–4]. It has found applications in economics and social sciences [5,6]. It has a fascinating history starting from Brown and Bachelier and from Rayleigh to Einstein. A review of the history of this field can be found in [7]. In a standard and well-studied formulation of random walk, the walker is unbiased and has no memory. The steps are randomly chosen in one of the allowed directions. We consider a variant in which he remembers the past and keeps track of total visits to the neighboring site. This could have relevance in search or foraging where one may choose to revisit or avoid previously visited sites. An extreme version in which it is not allowed to visit previously visited Site is known as self-avoiding walk. It is a difficult problem and has resisted all attempts of exact analytic solution. However, results from self-avoiding walks have found several applications, most noteworthy applications being in polymer chemistry. There have been previous attempts in this direction. In particular, exponential model has been studied in some works. All walks are enumerated and each walk has a weight $\exp(-g\sum n_i^\alpha)$ where $n_i$ is occupation number of the site [8]. It has been extended to a power-law case [9]. However, we cannot study very long walks with exact enumeration and we have carried out Monte-Carlo simulations for studying long walks with memory.

## 2. The Model

The model described below can be simulated to Cartesian topologies in any dimensions and even complex networks. However, in this work, we study a walk on 1-d lattice. Consider a one-dimensional array of length N. We associate variable $v(i)$ to each site i, $1 \leq i \leq N$. The variable $v(i)$ denotes number of visits to a given site i by the walker. The probability of visiting a particular site i depends on the number of visits $v(i)$ to that site. This probability depends on a $f(v(i))$ where f is some function of number of visits. The walker starts at some point k. Now it can jump to site k+1 or k-1. We consider periodic boundary conditions. We associate the following probabilities to jump to nearest neighbors.

i) Probability of visiting site k+1 as $f(v(k+1))/(f(v(k+1))+f(v(k-1)))$ and
ii) The probability of visiting site k-1 as $f(v(k-1))/(f(v(k+1))+f(v(k-1)))$
Where f is some function. We classify the walk into two classes.

a) Tourist walk: The site is more attractive if it is less visited previously. The function $f(v)$ is a monotonically decreasing function of v.
b) Avoiding walk: The site is less attractive if it is less visited in past. The function $f(v)$ is a monotonically increasing function of v. We consider following functional forms
i) $f(v)=P_\alpha(v)=(v+1)^\alpha$
ii) $f(v)=E_r(v)=\exp(rv)$
For $P_\alpha(v)$, we have added +1 so that probability is well defined for unvisited sites for $\alpha<0$. The walk is avoiding if $\alpha>0$ or $r>0$. The walk is tourist walk if $\alpha<0$ or $r<0$. The cases $\alpha=0$ or $r=0$, correspond to normal random walk.

## 3. Simulation and Results

Now several interesting questions can be asked and quantifiers can be defined.

### 3.1. Number of distinct sites visited

Let $d_N(t)$ be the number of distinct sites visited at time $t < t_{max}$ for both $P_\alpha(v)$ and $E_r(v)$ with $N> 2 t_{max}$, so that the finite size effects do not affect results.

1) We compute $d_N(t)$ for $P_\alpha(v)$ for $t_{max} = 10^8$. We study $\alpha$=-4, -3, -2, -1 on negative side. And $\alpha$= 0.2, 0.3, 0.4, 0.5, 0.6, 1.0 on positive side. We also plot $\alpha=0$ for comparison with normal random walk.

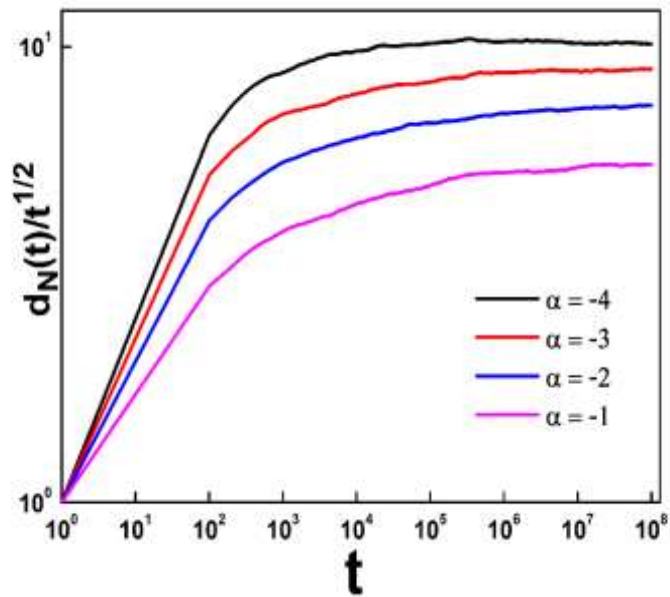

**Figure 1.** Plot for $d_N(t)/t^{1/2}$ as a function of t for $\alpha$=-4,-3,-2,-1.

i) For α<0, $d_N(t)$ grows faster than $t^{1/2}$ initially. But eventually $d_N(t)$ grows as $t^{1/2}$. (See Fig. 1). Thus the asymptotic behavior of α<0 is same as for α=0 which is a normal random walk without memory.

ii) For 0.5>α>0, $d_N(t)$ grows slower than $t^{1/2}$ and distinct power law is obtained.

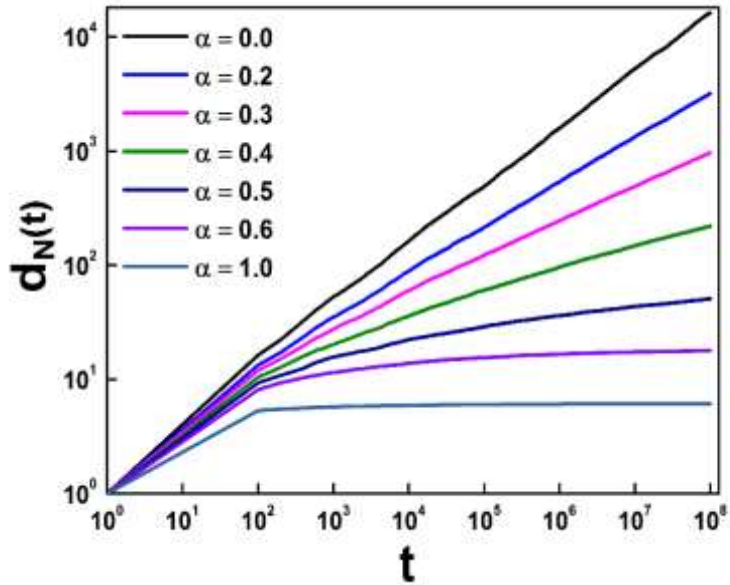

**Figure 2.** Plot for $d_N(t)$ vs t for α = 0, 0.2, 0.3, 0.4, 0.5, 0.6, 1.

iii) For α close to 0.5, distinct behavior is obtained, the exponent goes to zero as α→0.5. We observe that $d_N(t) \sim \log(t)$ for α = 0.5 (see fig. 3). This growth is slower than any power-law.

iv) For α>0.5, it is even slower than logarithmic and particle is confined to a small region. There is no increase in number of covered sites asymptotically.

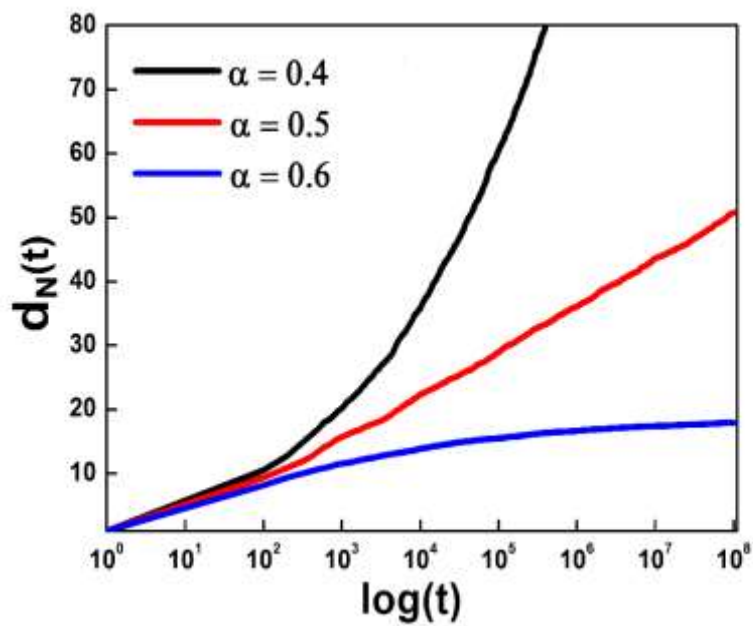

**Figure 3.** The plot for $d_N(t) \sim \log(t)$ showing logarithmic behavior for α=0.5

2) We compute the same quantity for exponential function. We compute $d_N(t)$ for $E_r(v)$ for r = -1. In this case, even asymptotically, a different power is obtained. The power is 2/3. This behavior is shown in Fig, 4.

For r = 1, $d_N(t)$ saturates and does not exceed 5 - 6 for t = $10^8$.

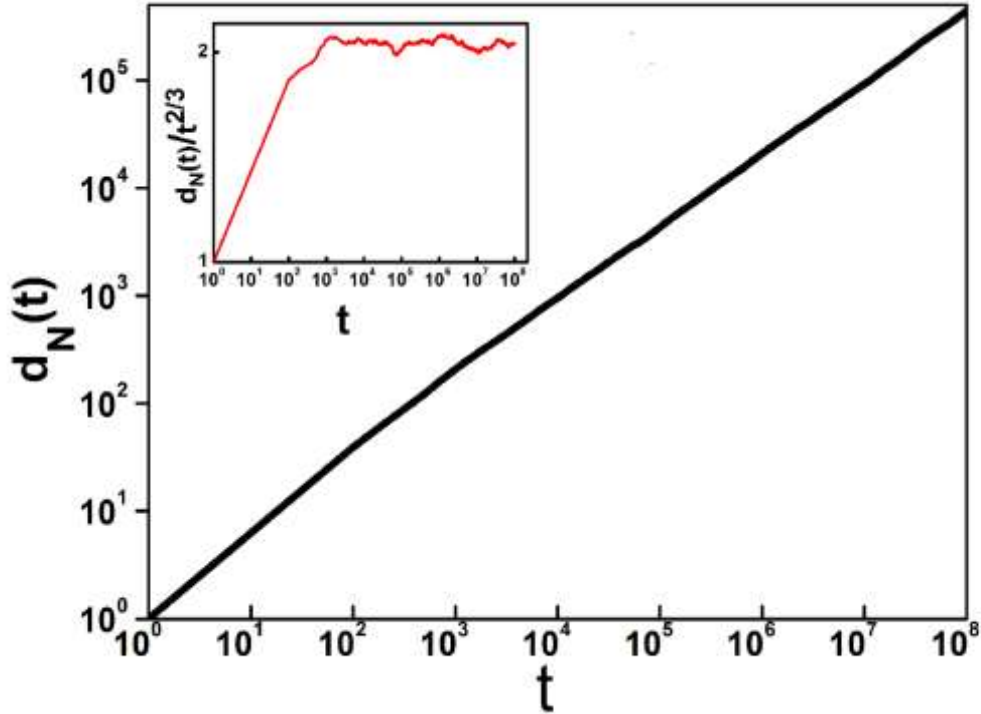

**Figure 4.** The main figure: the number of distinct site visited is plotted as a function of t. Inset: $d_N(t)/t^{2/3}$ plotted vs time.

*3.2. Lattice covering time*

Let $T_N$ be the time at which all the sites in the lattice are visited at least once. We study the distribution $P(T_N)$ of $T_N$ as well as average $<T_N>$ as a function of N for various forms of function f(v).

1) We compute $T_N$ for $P_\alpha(v)$ for α= -2, -1, -0.5, 0, 0.1, 0.2, .3.

i>For α<0, $d_N(t)$ grows faster than $t^{1/2}$ initially, but the growth slows down and eventually, it grows like $t^{1/2}$ asymptotically. Thus asymptotic properties for α<0 are the same as α=0 i.e. the normal random walk. This is very similar to the behavior of $d_N(t)$ studied in above section. Thus, $T_N \propto N^z$ with z=2, for α ≤ 0.

ii>The behavior changes for α>0 and z >2 for α>0.

The behavior for positive and negative values of α is shown in Fig. 5. We have also plotted the case α=0 for reference. It is clear that asymptotic behavior of this quantity is similar to behavior of simple random walk for α<0.

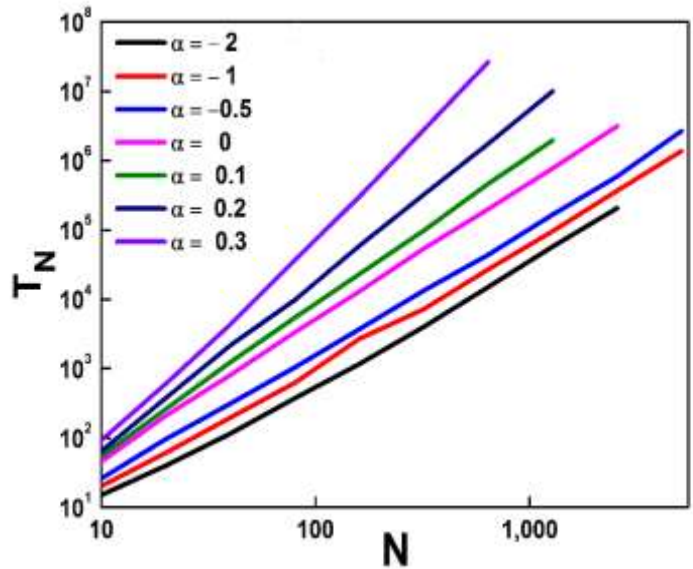

**Figure 5.** Plot for $T_N$ as function of N for $\alpha$= -2, -1, -0.5, 0, 0.1, 0.2, 0.3 for $P_\alpha(v)$. The slope does not change for $\alpha \leq 0$ (Bottom 4 lines). But there is a discernible change $\alpha > 0$.

2) We compute $T_N$ for $E_r(v)$ for r = -1. In this case power is found to be 3/2. Thus we observe that for $E_r(v)$, $z \sim 3/2$ (see fig. 6)

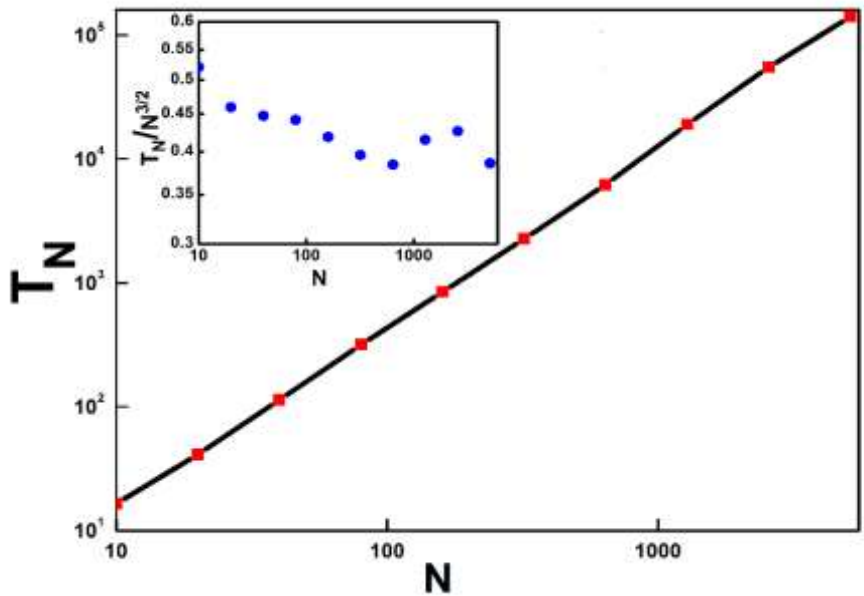

**Figure 6.** Plot for $T_N$ vs N for $E_r(v)$. Inset: The plot of $T_N/N^{3/2}$ vs N.

### 3.3. Density profile at various times

Let $D(i,t) = v(i)/t$. It shows how often a site i is visited. We study the density profile at different times for various form of function f(v). We exclude initial position of walker at time t=0 in this profile.

1) Consider $\alpha > 0$. In Fig. 7, we have plotted density profiles for various time including $t=10^2$, $t=10^3$, $t=10^4$, for $\alpha > 0$ as a function of site i. We observe that, for both $\alpha=0.1$ and $\alpha=0.2$ shows a single peak at i=0.

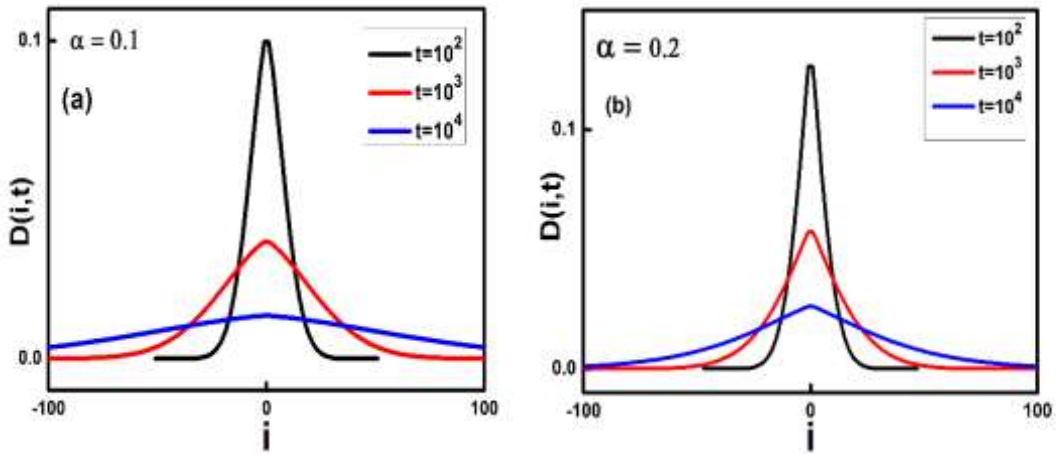

**Figure 7.** The D(i,t) plotted at various time including $t=10^2$, $t=10^3$, $t=10^4$, as a function of site i. a) α=0.1, b) α=0.2

Consider α<0. We have plotted density profiles for various time including $t=10^2$, $t=10^3$, $t=10^4$, for α<0 as a function of site i. Here we observe that, for α= -1, α= -2, α= -3, α= -4 density profile shows a two peaks. The peaks go further as time proceeds. This behavior is shown in Fig. 8.

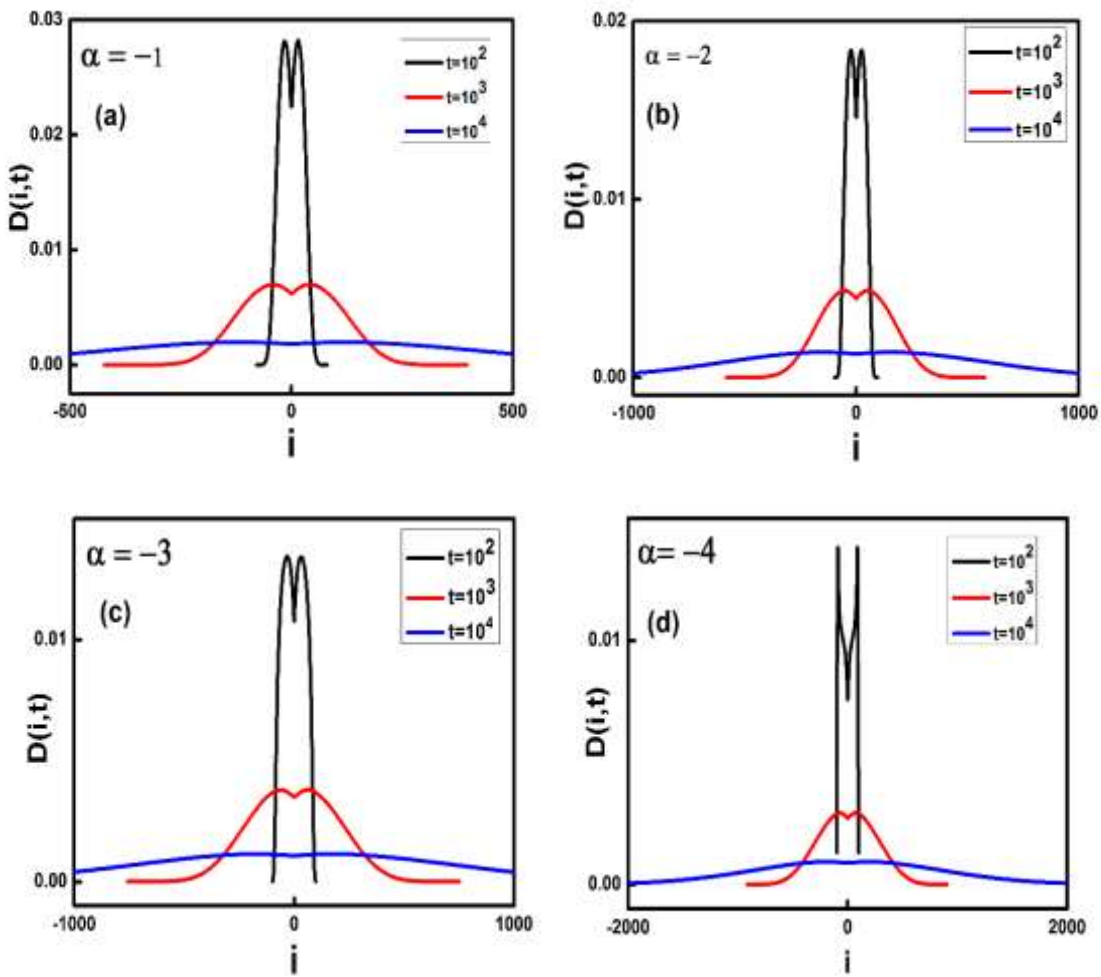

**Figure 8.** The D(i,t) plotted at various time including $t=10^2$, $t=10^3$, $t=10^4$, as a function of site i, a) α=-1, b) α=-2, c) α=-3, d) α=-4.

2) For E(v)=exp[-v(i)], we observe behavior similar to α<0. There are two symmetrically placed peaks in density profile. The absolute value at which the peaks occur is larger at longer times. This behavior is similar to α<0 described above.

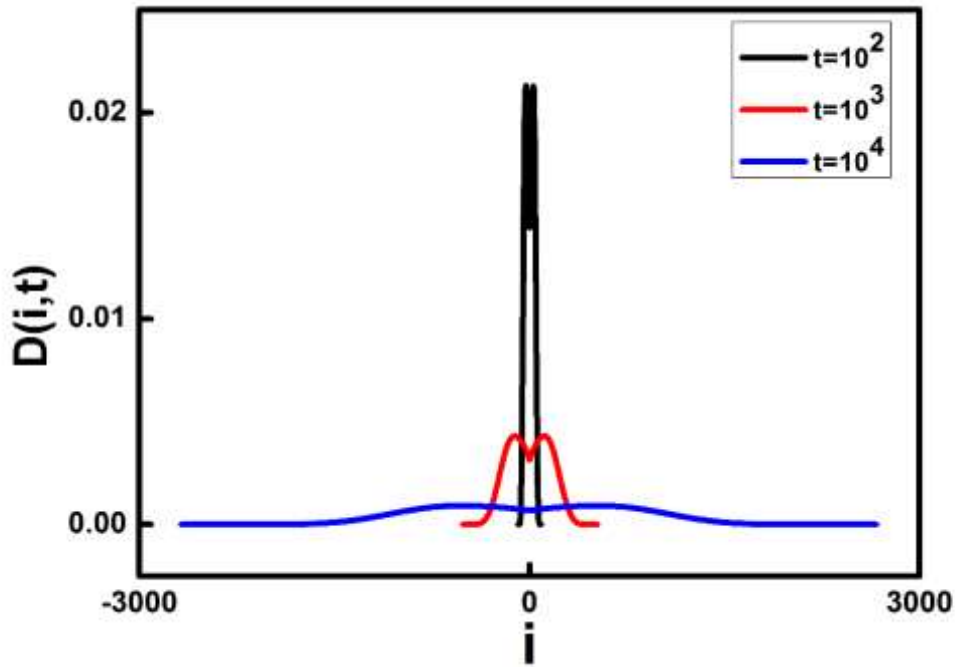

**Figure 9.** Plot for D(i,t) vs i at various time including t=$10^2$, t=$10^3$, t=$10^4$, For $E_r$(v)=exp[-v(i)].

**Acknowledgement**
PMG thanks DST-SERB (EMR/2016/006685) for financial assistance. MCW thanks Council of Scientific and Industrial Research (C.S.I.R.), JRF (09/128(0097)/2019-EMR-I).